%% file: dirsch.tex
\documentclass[preprint,11]{aastex}

\usepackage{verbatim,float}
\usepackage{times}
\usepackage{amsmath}
\usepackage{graphics}
\usepackage{epsfig}
\usepackage{psfig}
\usepackage{longtable}

\begin{document}

\title{The Globular Cluster System of NGC 1399: III. VLT Spectroscopy and
Database\altaffilmark{10}}

\shorttitle{Spectroscopy of NGC 1399 Globular Clusters \sc{iii}}
\shortauthors{Dirsch et al.}

\author{B. Dirsch \altaffilmark{1}, T. Richtler \altaffilmark{1}, D. Geisler
 \altaffilmark{1}, K. Gebhardt \altaffilmark{2}, M. Hilker \altaffilmark{3},
M.V. Alonso \altaffilmark{7}, J.C. Forte \altaffilmark{6}, E.K. Grebel \altaffilmark{8},
L. Infante \altaffilmark{4}, S. Larsen \altaffilmark{5}, D. Minniti \altaffilmark{4}, M. Rejkuba \altaffilmark{5}}

\altaffiltext{1}{Universidad de Concepci\'on, Departamento de F\'{\i}sica,
                Casilla 160-C, Concepci\'on, Chile}
\altaffiltext{2}{Department of Astronomy, University of Texas at Austin,
TX78712, Austin, USA}
\altaffiltext{3}{Sternwarte der Universit\"at Bonn, Auf dem H\"ugel 71, 53121 Bonn, Germany}
\altaffiltext{4}{Departamento de Astronom\'{\i}a y Astrof\'{\i}sica, P. Universidad Cat\'olica, Vicu\~na Mackenna 4860, Santiago 22, Chile}
\altaffiltext{5}{European Southern Observatory, Karl-Schwarzschild-Str.2, D-85748 Garching,Germany}
\altaffiltext{6}{Universidad Nacional de La Plata, Facultad Ciencias Astronomicas y Geofisicas,
Paseo del Bosque S/N, 1900-La Plata, UNLP and CONICET, Argentina}
\altaffiltext{7}{Observatorio Astron\'omico de
C\'ordoba,  Laprida  854, C\'ordoba, 5000, CONICET, Argentina\\
CNRS UMR 5572, Observatoire Midi-Pyr\'en\'ees, 14 Avenue E. Belin, 31400
Toulouse, France}
\altaffiltext{8}{Max-Planck Institut f\"ur Astronomie, K\"onigstuhl 17, D-69117, Heidelberg, Germany}
\email{bdirsch@cepheid.cfm.udec.cl}

\altaffiltext{10}{Based on observations collected at the European Southern
Observatory, Cerro Paranal, Chile; ESO program 66.B-0393.}


\begin{abstract}

Radial velocities of 468 globular clusters around NGC\,1399, the central galaxy
in the Fornax cluster, have been obtained with FORS2 and  the Mask Exchange Unit (MXU) at
the ESO Very Large Telescope.  This is the largest sample of globular cluster
velocities around any galaxy obtained so far. The mean velocity uncertainty is
50\,km/sec. This data sample is accurate and large enough to be used in studies
of the mass distribution of NGC\,1399 and the properties of its globular cluster
system. Here we describe the observations, the reduction procedure, and discuss
the uncertainties of the resulting velocities. The complete sample of cluster
velocities which is used in a dynamical study of NGC 1399 is tabulated. A
subsample is compared with previously published values.

\end{abstract}


\section{Introduction}

In the context of understanding the distribution of dark matter in galaxies,
early-type galaxies in the centers of galaxy clusters play an important role. In
these places, the highest dark matter densities are found and accordingly, the
transition between luminous and dark matter dominated regions can be best
investigated. In the inner regions of elliptical galaxies this can be done by
long slit spectroscopy measuring integrated spectral line profiles (see
Kronawitter et al. (\cite{kronawitter00}) and Gerhard et al. (\cite{gerhard01})
for a large sample of galaxies).

In the outer region of an elliptical galaxy -- in an area where dark matter
begins to dominate -- this approach is hardly feasible due to the low surface
brightness. Globular clusters (GCs) and planetary nebulae (PNe), which can be
traced to large radii, are the best tools to study the dynamics at larger radii.
However, the distribution of radial velocities of particles is degenerate with
respect to the galaxy mass profile and the orbital structure of the dynamical
probes (e.g. Merritt \& Tremblay  \cite{merritt93a}).
If no other constraints on the velocity dispersion tensor and mass
distribution are available, then radial velocities for several hundred or
more clusters are required to break this degeneracy (e.g., Merritt \cite{merritt93}).
Therefore, if we want to lift this degeneracy, we are limited to those nearby large
elliptical galaxies with sufficiently populous cluster systems. M\,87 is such a
candidate in the northern hemisphere.  Cohen \& Ryzhov \cite{cohen97} observed
about 200 radial velocities, using the Keck telescope. This sample has been
improved and enlarged to about 350 velocities (C\^ot\'e et al. \cite{cote01},
Hanes et al. \cite{hanes01}).

Another promising target is the central galaxy of the Fornax cluster, NGC 1399.
It's populous globular cluster system (GCS) has been
investigated by Hanes\,\&\,Harris (\cite{hanes86}), Bridges et al.
(\cite{bridges91}), Wagner et al. (\cite{wagner91}), Ostrov et al.
(\cite{ostrov93}), Kissler-Patig et al. (\cite{kisslerpatig97}), Grillmair et al.
(\cite{grillmair99}) and most recently by Dirsch et al. (\cite{dirsch03},
Paper\,{\sc i}) the latter covering a field of $36\arcmin\times36\arcmin$. The GCS has been
the target of three spectroscopic studies. Grillmair et al. (\cite{grillmair94})
obtained low-dispersion spectra of 47 GCs within a radial distance of 9\arcmin \
around NGC\,1399 with the Low Dispersion Survey Spectrograph on the AAT. A
typical uncertainty of their velocities is 150\,km/sec. Another contribution
comes from Minniti et al. (\cite{minniti98}) who used the ESO Multi-Mode
Instrument at the ESO 3.5\,m New Technology Telescope to measure velocities of 19
GCs. Their velocity uncertainties are also large: the mean uncertainty is 128\,km/sec.
The velocities with the smallest errors available so far are obtained by
Kissler-Patig et al. (\cite{kisslerpatig98}) with the Low Resolution Imaging
Spectrograph at the Keck telescope. They determined the velocities of 21 GCs with
an average uncertainty of 45\,km/sec. Kissler-Patig et al.
(\cite{kisslerpatig99}) compiled the velocity determinations from these three
studies, ending up with 74 GC velocities.

Beside the GCs, planetary nebulae (PNe) have been used for dynamical studies of
NGC\,1399 by Arnaboldi et al. (\cite{arnaboldi94}) and Napolitano et al.
(\cite{napolitano02}). Both studies used the same database of 37 PN velocities in
the inner 4\arcmin \ of NGC 1399.  At smaller radial distances ($\le 97\arcsec$)
the galaxy light itself has been employed to constrain dynamical models by Saglia
et al. (\cite{saglia00}), Kronawitter et al. (\cite{kronawitter00}), Gerhard et
al. (\cite{gerhard01}).

In order to greatly improve our knowledge of its GCS kinematics and galaxy
halo dynamics,
we used the FORS2 at the VLT to obtain spectra of globular clusters out to a
radial distance of 8\arcmin \, which corresponds to approximately 45\,kpc.
A distance modulus of $\mathrm{m}-\mathrm{M}=31.4$ (19\,Mpc) is used
(see Dirsch et al. \cite{dirsch03} and discussion therein).

The observations aim at a significant improvement in the number and quality of
dynamical probes.
The selected cluster candidates have luminosities brighter than
$\mathrm{V}=22.5$\,mag.  An ideal telescope to obtain the required spectra of
such faint objects in the southern hemisphere is the ESO
Very Large Telescope. A spectroscopic resolution of approximately 3\AA \ is
desirable for our task and thus FORS2 with the MXU ({\bf M}ask e{\bf X}change {\bf
U}nit)  is an appropriate instrument since it allows the simultaneous observation of
some 100 objects.

In this article we present the data analysis and error discussion of our
measurements. The complete dataset of 502 cluster velocities is presented which
is the basis of the dynamical analysis published in Paper\,{\sc ii}. These
papers are part of a bigger effort to study the dynamics and stellar content of
NGC 1399, which also includes new photometric wide field observations published
in Paper\,{\sc i}.

\section{Spectroscopic mask preparation}

The preparation of the slit masks we used for observing NGC1399 at the VLT
is an important step in the overall investigation. Hence, we describe this
first step in detail, before discussing the observations themselves.
Our candidates GCs
 have been selected on the basis of new wide-field photometry in
Washington C and Kron-Cousins R of the cluster system of NGC 1399, obtained at the CTIO
4m MOSAIC system. For a description of the photometric observations
the reader is referred to Paper\,{\sc i}.

For the point source selection we used a color criterion
($0.9<(\mathrm{C-T1})<2.4$) and also a morphological criterion based on the
SExtractor class (Bertin \& Arnout \cite{bertin96}; only objects with
{\it class} larger than 0.8 are retained,
based on artificial star experiments and comparison with the VLT preimages. This
morphological selection limited our sample to objects brighter than R=23.3\,mag
which allows us to obtain the required signal-to-noise in the spectra.

For accurate slit placement on the MXU masks, preimaging is required. Service
mode observations with FORS2 at the VLT in October 2000 provided us with four
FORS2 fields, each of them $7\arcmin\times7\arcmin$, arranged in a quadratic
configuration (see Fig.1). Each field has been exposed for 30\,sec in Johnson V.
We identified the selected clusters in these images and placed the slits
accordingly. Since the VLT images had superior seeing (around $0\farcs6$),
compared to our MOSAIC observations, we used them to check the star-like
appearance of the selected objects.  This left us with about 500 globular cluster
candidates.

A major point during the preparation of the masks , which was done with the ESO
FORS Instrumental Mask Simulator (FIMS
\footnote{http://http.hq.eso.org/observing/p2pp/OSS/FIMS/FIMS-tool.html}), were
the choices of the slit width and particularly the slit lengths. We decided for
a slit width of 1$\arcsec$, which turned out to be appropriate for the normally
sub-arcsecond seeing conditions.

The slit length was a more complicated issue. The spectra of our fainter targets
are sky-dominated and the sky subtraction is best done by measuring object and
sky in the same slit. However, the severe disadvantage with slits of the
appropriate length, 5$\arcsec$ or longer, is that they would cover up many objects
which otherwise would have been observable with smaller length, contrary to our
objective of observing as many targets as possible. Moreover, it considerably
lowers the flexibility of choosing slit positions according to the random
location of targets with a high surface number density in a mask. Therefore, we
decided to set sky slits independently from the object slits and to subtract the
sky after the wavelength calibration, trusting in its accuracy.

For most slits, the size then was $1\arcsec\times 2\arcsec$, being
$1\arcsec\times5\arcsec$ only for the few bright objects (normally foreground
stars), which were needed to accurately position the mask. In this manner, we
could fill a mask with up to 120 slits, typically 40 cluster candidates, 40 sky
slits, and many miscellaneous objects, among them other point sources not
matching our strict selection criteria, or background galaxies.

The color-magnitude diagram of the final sample for which we obtained velocities
is shown in Fig.\,\ref{fig:spec_cmd} together with its color distribution that is
compared to the color distribution of the total MOSAIC sample within the same
radial range.  This figure illustrates that the final sample is not subject to
color selection effects.

\section{The observations}

\begin{figure}[t]
\centerline{\resizebox{\hsize}{!}{\includegraphics{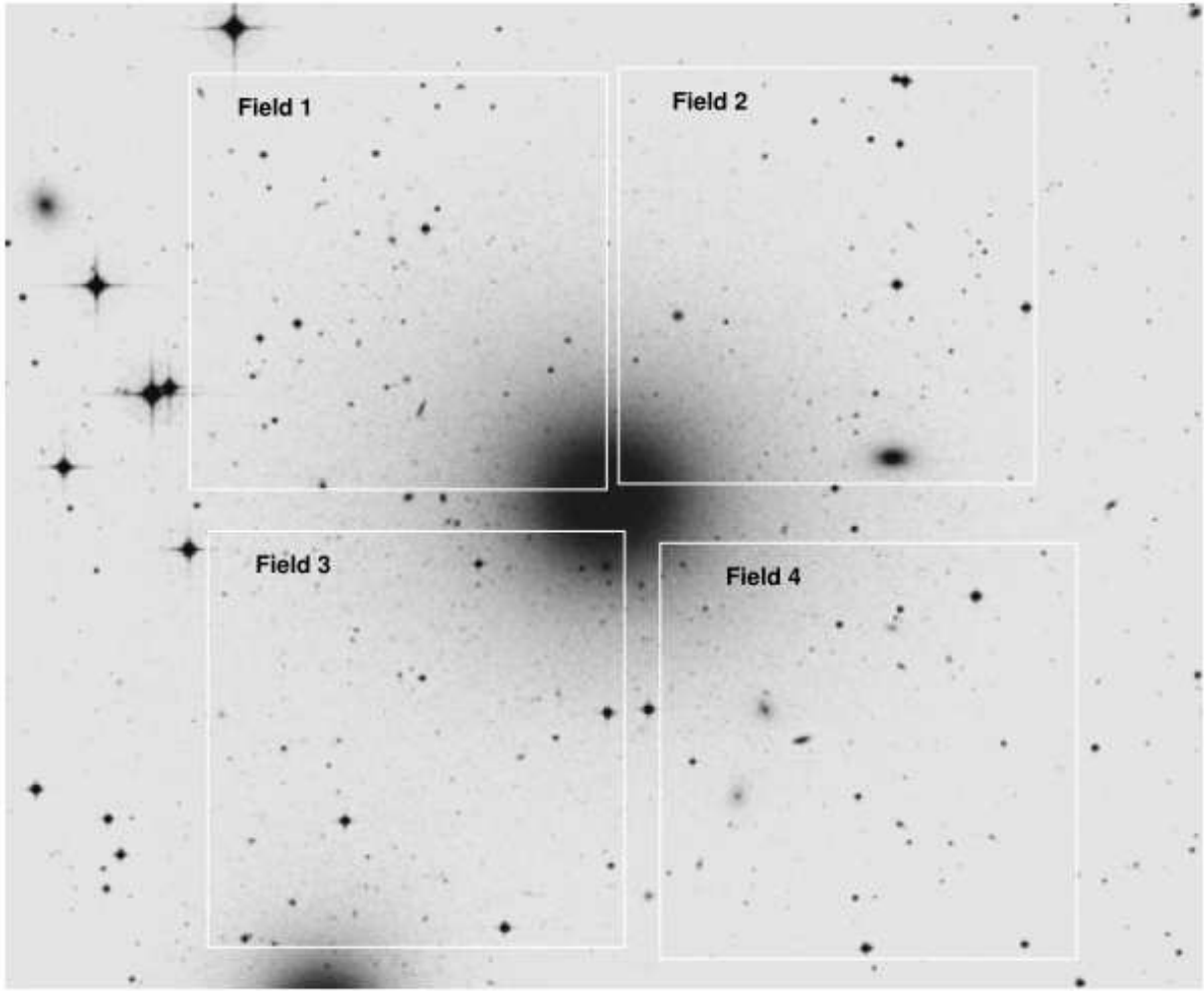}}}
\caption{The positions of the four fields (each having a size of $7\arcmin\times7\arcmin$
	in which we obtained the spectra
	are overlayed on a DSS image (blue filter). East is to the left, North
	is to the top, the size of the image is $19\farcm3\times15\farcm6$. }
\label{fig:masks}
\end{figure}

The observations have been performed during 30.11.--2.12.2000 at the European
Southern Observatory at Cerro Paranal with the Very Large Telescope facility. The
telescope was UT2 (Kueyen) and the instrument the focal reducer FORS2, equipped
with the MXU.  The detector was a SiTE SI-424A backside thinned CCD with a pixel
size of 24$\times$24 microns. The total field of view is $6\farcm83\times
6\farcm83$. The observations are summarized in Table\,\ref{tab:observations}. The
four VLT fields in which the masks were placed are shown overlayed on a DSS image
of NGC\,1399 in Fig.\,\ref{fig:masks}.

We used the 600B grism without filter, which provides a resolution of
around 2.5\AA. The spectral range covered was about 2000
{\AA}. However, due to the position of a given slit in the mask, the limiting
short- and long wavelengths varied considerably, ranging from 3500 {\AA} to 6500
{\AA}. In most cases, the grism efficiency degraded the signal-to-noise
short-wards of 3800 {\AA} drastically, so this region could not be used. Mask 89
has been observed using the 300V grism with a lower spectral resolution
(110\AA/mm). This has been done to see the effect of using a certain grism. The
analysis showed no difference in the quality of the derived velocities for the
two grisms.

We exposed 13 masks (exposure times were either 45 min or 2$\times$45 min). The
seeing was always sub-arcsecond and ranged from $0\farcs6$ to $0\farcs9$.
Flat-fielding was done with standard lamp flats. For the wavelength calibration,
a HeHgCd lamp was exposed during service day-time calibration.

The total number of spectra obtained is 1462. This sample is composed of 531 sky
spectra (see Table\,\ref{tab:efficiency}), 512 spectra of cluster candidates, 190
spectra of point sources of unknown nature at the time of mask preparation
(stars, unresolved galaxies or globular clusters), 176 galaxies, and 53
``bright'' objects, mainly stars which were needed to adjust the mask. Since some
objects (about 80) have been observed in two different masks in order to assess
the velocity uncertainties, the total number of objects is smaller by this number
than the number of spectra. There are two reasons why we found clusters in the
(randomly selected) point source sample: firstly, the MOSAIC data is not dithered
and thus for a considerable fraction of clusters no photometry was obtained.
Secondly, our employed morphological selection criteria were very strict and the
completeness of faint cluster identification was rather low (see also Paper {\sc
i}).

Table \ref{tab:observations} lists the relevant data of the observations,
starting with the number of the respective FORS field, the mask number, the
center coordinates (J2000), the exposure time (90 min means that two exposures of
45 min have been stacked).

\section{The data reduction}

The basic reduction (bias subtraction, flatfielding, trimming)  has been done
with standard procedures within IRAF. For the later extraction of the spectra the
{\it apextract} package has been used.

Regarding the removal of cosmic rays, we found after some experimenting that the
task FILTER/COSMICs under MIDAS gave the most satisfactory results. A few
artefacts remained in spite of that, best visible in the faint spectra.

In the flat-fielded image the spectra have a separation of only a few pixels and
are curved along the dispersion axis which has to be fitted before they can be
extracted. The curvature is strongest at the frame edges, where the deviation
from the center to the edge is approximately 6\,pixels. We traced each spectrum
along the dispersion axis on the flatfield image of the mask because of their
much clearer signal.  We then kept the tracing parameters and optimized the size
of the apertures with the goal to minimize the sky contribution in the object
spectra which has been done on the science spectra. The final sizes of the
apertures depend on the seeing and varies between $1\arcsec$ and $1\farcs4$. On
some masks a few spectra overlap which have been excluded from further analysis.

We employed the IRAF task {\it identify} to calibrate both object and sky
spectra. Typically around 18 He, Hg and Cd-lines were kept for the line-list.
The calibration uncertainty is $\pm0.04$\,\AA.

The bright O\,{\sc i} skyline at 5567\,\AA \ which is present in all spectra can
be used to correct zero point differences between the masks.  Differences up to
$\pm 0.8$\,\AA \ have been found. The reason for these systematic differences is
the uncertainty in the mask placement which is cited in the MXU manual
\footnote{http://www.eso.org/instruments/fors/userman/} to be 13\,microns.  This
is approximately half a pixel or 0.6\AA \ using the 600B grism and thus in good
agreement with our measured zero point differences.

Within a mask the position of the O\,{\sc i} sky line was constant to within
0.04\AA \ consistent with the calibration uncertainty. An exception is mask \#89
(with the grism 300V), where we found a systematic linear increase/decrease of
the O\,{\sc i} line position from 5566.3\,\AA \ to 5567.91\,\AA. We corrected for
this with a linear interpolation in the wavelength calibration. The systematic
behavior in mask\,\#89 is probably due to a slight rotation of the mask.

For the sky subtraction we used spectra that were observed through slits placed
in empty sky areas. For each object adjacent sky spectra within a certain radial
distance from NGC\,1399 have been selected to attempt to obtain the best sky spectrum. The
radial width of these annuli had been chosen to be smaller near to NGC\,1399 and
wider further away ($1\farcm2$ for distances smaller than $2\farcm5$ and
$1\farcm7$ for larger distances ).  Typically 3 to 11 sky spectra were averaged
and subtracted from the object spectra. The quality of the subtraction was judged
by examining the residuals around the bright sky lines in the red.

\begin{figure*}[t]
\centerline{\resizebox{12cm}{!}{\includegraphics{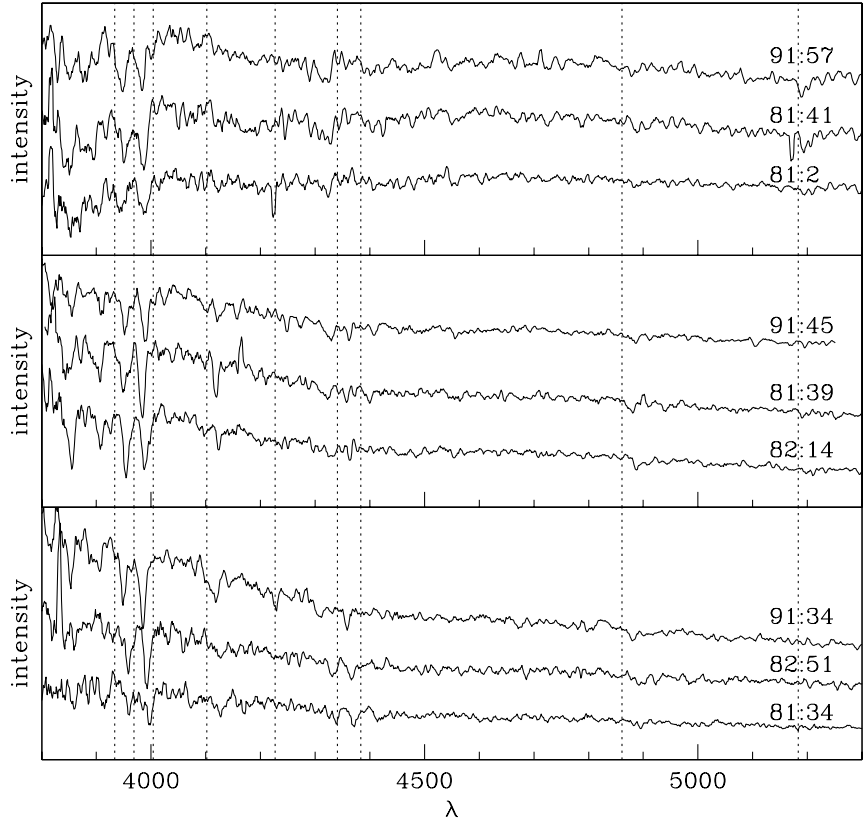}}}
\caption{Example spectra of brighter cluster candidates (smoothed by 3.6 \AA \ for
	illustration) in three color regimes -- {\bf upper panel}:
	$\mathrm{C-T1}\approx1.95$, {\bf middle panel}: $\mathrm{C-T1}\approx1.35$,
	{\bf lower panel}: $\mathrm{C-T1}\approx1.15$. We
	indicated the rest frame wavelengths of some prominent absorption
	features by the dotted lines. From left to right these are: Ca\,{\sc k}
	(3933.66\AA), Ca\,{\sc h} (3968.47\AA), Fe\,{\sc i} (4045.81\AA),
	H\,$\delta$ (4101.73\AA), Ca\,{\sc i}\,\&\,Fe\,{\sc i} blend (4226.73\AA),
	H\,$\gamma$ (4340.46\AA), Fe\,{\sc i} (4383.54\AA), H\,$\beta$ (4861.32\AA),
	Mg\,{\sc i} (5183.64\AA). The number at the right side of the spectra is
	the identifier of the objects explained in Sec.\,5.3 and used in
	Tab.\,3.}
\label{fig:spec_example}
\end{figure*}

This procedure resulted in good sky subtracted spectra as long as wavelengths
longer than 4000\AA \ are considered. For shorter wavelengths the uncertainties
in sensitivity (``flat-field uncertainties'')  between sky and object spectra
make the sky subtraction difficult. However, due to the low signal-to-noise of
the spectra blue-wards of 4000\AA \ this uncertainty is anyway not important.
Some example spectra around T1=21 are shown in Fig.\,\ref{fig:spec_example}

\section{Velocities}

\subsection{Velocity determination}

We determined velocities with two different techniques: first by using the
correlation with an object of known velocity and second by measuring the redshift
of an object directly with identified absorption lines.

The line measurements were performed with the help of the {\it rvidlines} task
within IRAF. Typically, around 15 features have been fitted for the velocity
determinations.

For the correlation technique we used the task {\it fxcor} implemented in IRAF
(the technique is described by Tonry\,\&\,Davis \cite{tonry79}).  We did not
utilize any Fourier filtering but we rather smoothed very noisy spectra with a
median filter (3 pixels).  The continuum has been subtracted by {\it fxcor} with
a spline-fit to the spectra and a $2-\sigma$ clipping algorithm around the fit
line. The range that has been proven to be best suited for the velocity
determination is 4500\AA \ to 5500\AA. However, for fainter objects we adjusted
the range individually to find the most significant correlation peak. As
template, we used a high S/N-spectrum (S/N about 40) of NGC\,1396, a small galaxy
with low intrinsic velocity dispersion and a spectrum similar to that of a
cluster. This object is on mask\,\#82, which accordingly served as the reference
mask. Zero-point differences with the other masks had been accounted for by using
the position of the sky O\,{\sc i}-line. We determined NGC\,1396's velocity by
measuring lines and obtained $815\pm8$\,km/sec as the heliocentric velocity.
This value is considerably lower than most older values from the literature:
da\,Costa et al. (\cite{dacosta98}) found $856 \pm 37$\,km/sec and in the RC3
catalogue $894 \pm 29$\,km/sec is given (de Vaucouleurs et al.
\cite{devaucouleur91}). However, Drinkwater et al.  (\cite{drinkwater01a}) quote
$808\pm8$\,km/s. The mean velocity of all clusters is 1438$\pm$15 km/s.
As for NGC 1399 itself, 
one finds 9 measurements of its radial velocity  with quoted uncertainties
consulting the NASA/IPAC Extragalactic Database. After skipping two of them 
(one has a discrepant value and the other a large
uncertainty of 200 km/s), the weighted mean value is 1442$\pm$9 km/s. Our mean radial
velocity of the entire cluster sample is 1438$\pm$15 km/s, so we are confident of
our absolute velocity calibration.

\subsection{Velocity uncertainties}

\begin{figure}[t]
\centerline{\resizebox{\hsize}{!}{\includegraphics{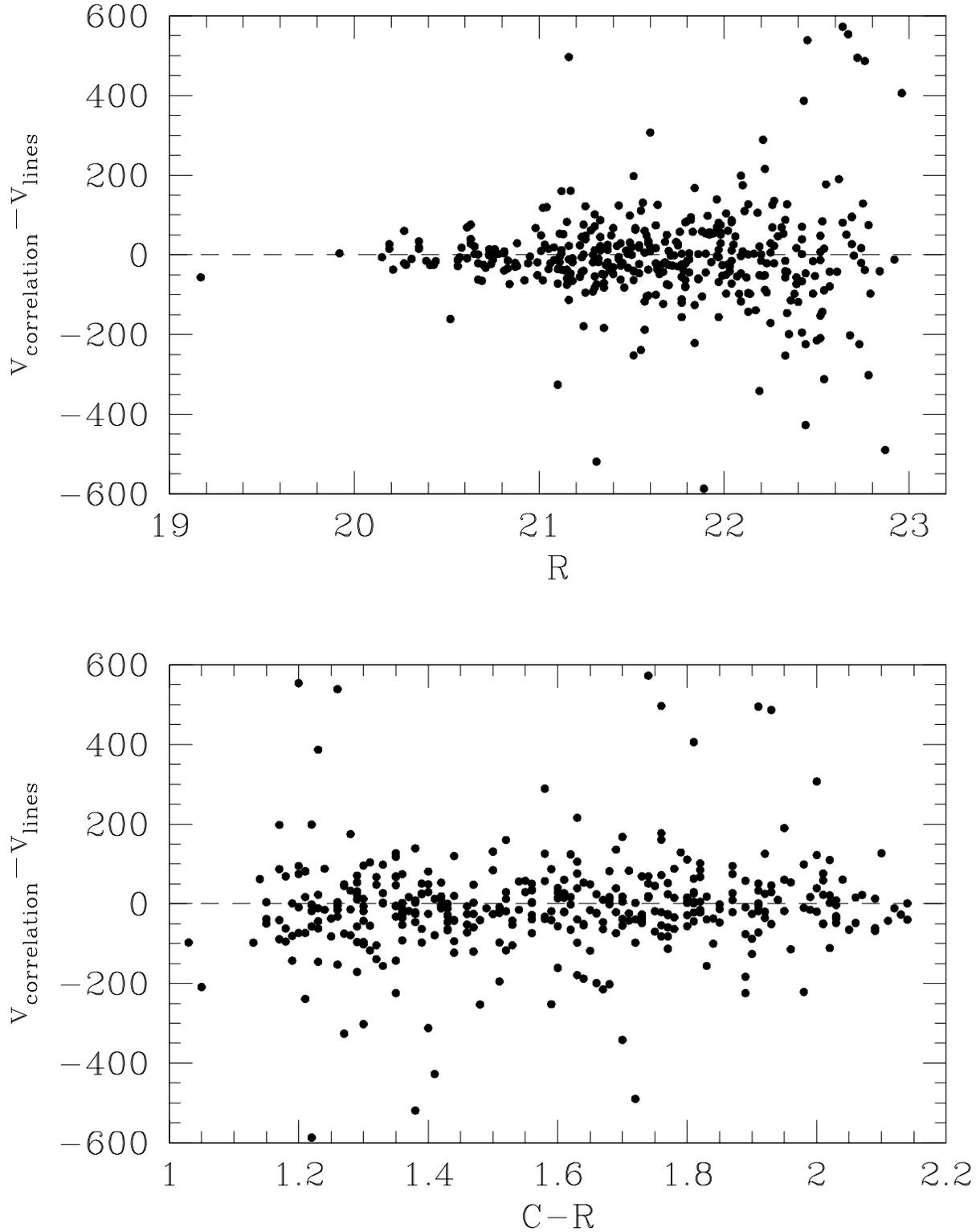}}}
\caption{The difference between the velocities determined via line measurements
	and the correlation technique are plotted in the {\bf upper panel} versus
	the brightness of the objects. It is expected and visible that for fainter
	objects the differences increase. In the {\bf lower panel} the differences
	are plotted versus their colors. No correlation with color can be seen.}
\label{fig:error1}
\end{figure}

The differences between cluster velocities that were derived with individual line
and with correlation measurements (shown in Fig.~\ref{fig:error1})  can be used to
study the velocity uncertainties. A dependence of the differences on the
brightness of the objects is expected and can be observed. However, deviations as
large as 500\,km/sec are more than 3-$\sigma$ larger than what is expected from
the measurement uncertainties. This indicates that at least for some measurements
systematic errors dominate over statistical errors. We will show later that
errors in the line measurements of noisy spectra are responsible for this
behavior.
The mean value of the difference between line and correlation measurements
is $-6.2\pm0.3$\,km/sec.  To derive the mean we
excluded clusters with absolute differences larger than 500\,km/sec.
In this case median and mean agree. The question remains whether this
is a significant deviation. The absolute scale of the correlation velocities are
based on the heliocentric velocity of NGC\,1396, for which we give an error of
8\,km/sec. Taking this into account the deviation from zero of the differences
is $-6\pm8$\,km/sec  and hence well inside the given uncertainty.

%

\begin{figure}[t]
\centerline{\resizebox{\hsize}{!}{\includegraphics{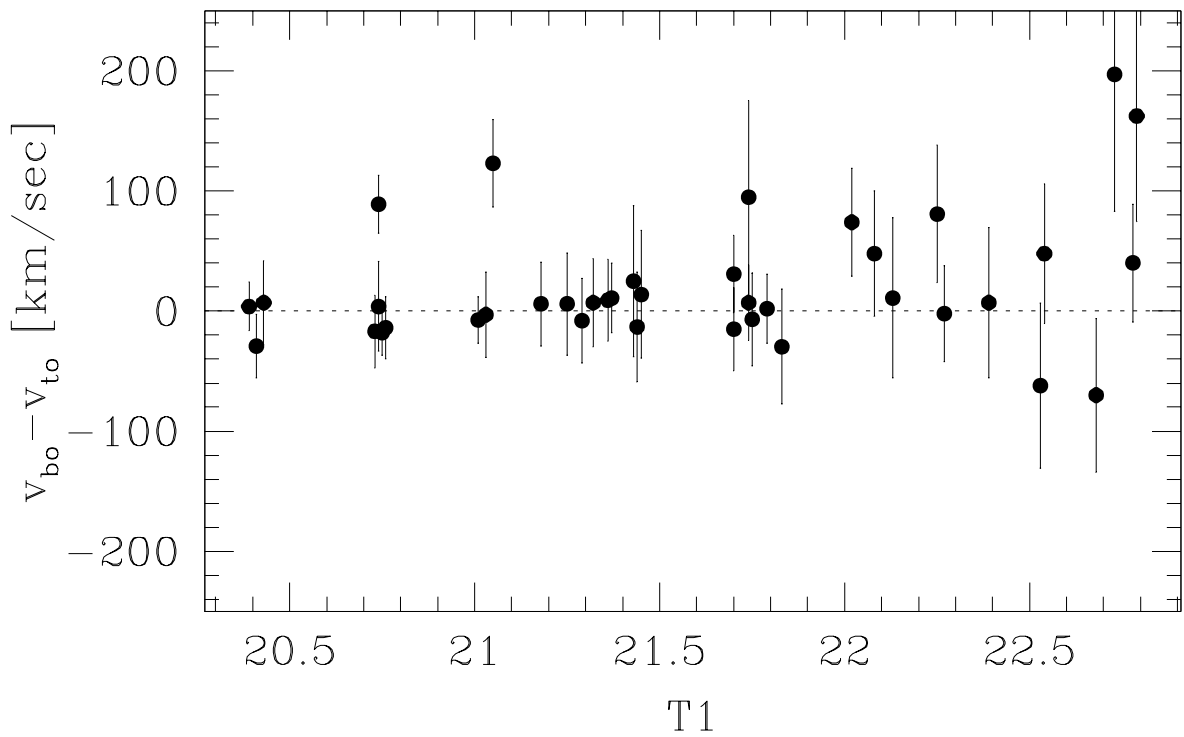}}}
\caption{Differences of the velocities determined on mask\,80 by line measurements by
	different authors, TR ($\mathrm{v}_\mathrm{to}$) and BD ($\mathrm{v}_\mathrm{bo}$).
	The scatter reflects the uncertainty resulting from independent treatment
	(tracing, wavelength calibration, line identification) of the data.}
\label{fig:error2}
\end{figure}

For mask\,\#80 the spectra extraction and subsequent velocity determination has
been done independently by two of us (TR and BD) and the differences for the line
measurements are plotted in Fig.~\ref{fig:error2}. The standard deviation between
the two measurements is $\sigma = 32$\,km/sec while we expect $37$\,km/sec from
the line measurement errors.  The scatter reflects the uncertainty resulting from
independent treatment (tracing, aperture definition, wavelength calibration, line
identification) of the data.

\begin{figure}[t]
\centerline{\resizebox{\hsize}{!}{\includegraphics{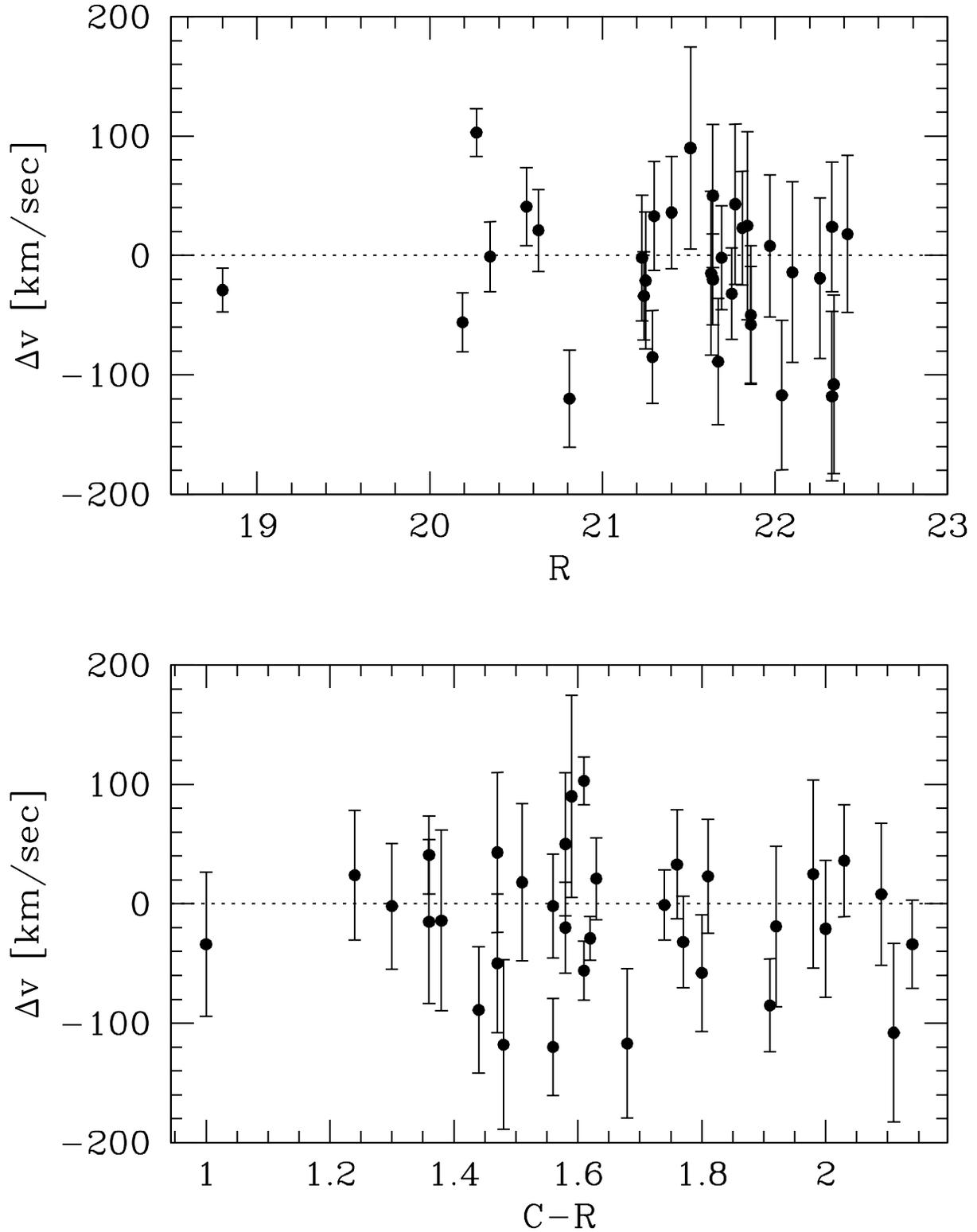}}}
\caption{Differences of the correlation velocities for common objects
	on different masks. No systematic difference can be seen.
	The scatter is compatible
	with the mean uncertainties of the velocities (see Sec.5.2).}
\label{fig:error3}
\end{figure}

The best estimation of the uncertainties is a comparison of measurements that are
obtained on different masks, which has been done for 31 point sources (29
clusters and two stars). The differences between the velocities are plotted in
Fig.\,\ref{fig:error3}, in dependence on brightness and color. The scatter of the
velocities measured via correlation is 57\,km/sec and in very good agreement with
the expectation from the individual uncertainties (52\,km/sec), which shows that
the given uncertainties are correct.
The scatter of the
velocities based on direct line measurements is 79\,km/sec and much larger than
expected from the individual errors (40\,km/sec).  This indicates that the errors
of the line measurements are dominated by systematic errors and are not of a
statistical nature. The reason for the systematic errors is most probably line
misidentification in noisy spectra. Line misidentification in noisy spectra also
accounts for small velocity uncertainties at relatively faint clusters.

Summarizing, we suggest to use the correlation velocities for any purpose, however,
we provide the line measurements for a consistency check. For this reason we do not
attempt to derive a more refined uncertainty estimation for the line velocities.

\subsection{The final sample}

\begin{figure}[t]
\centerline{\resizebox{\hsize}{!}{\includegraphics{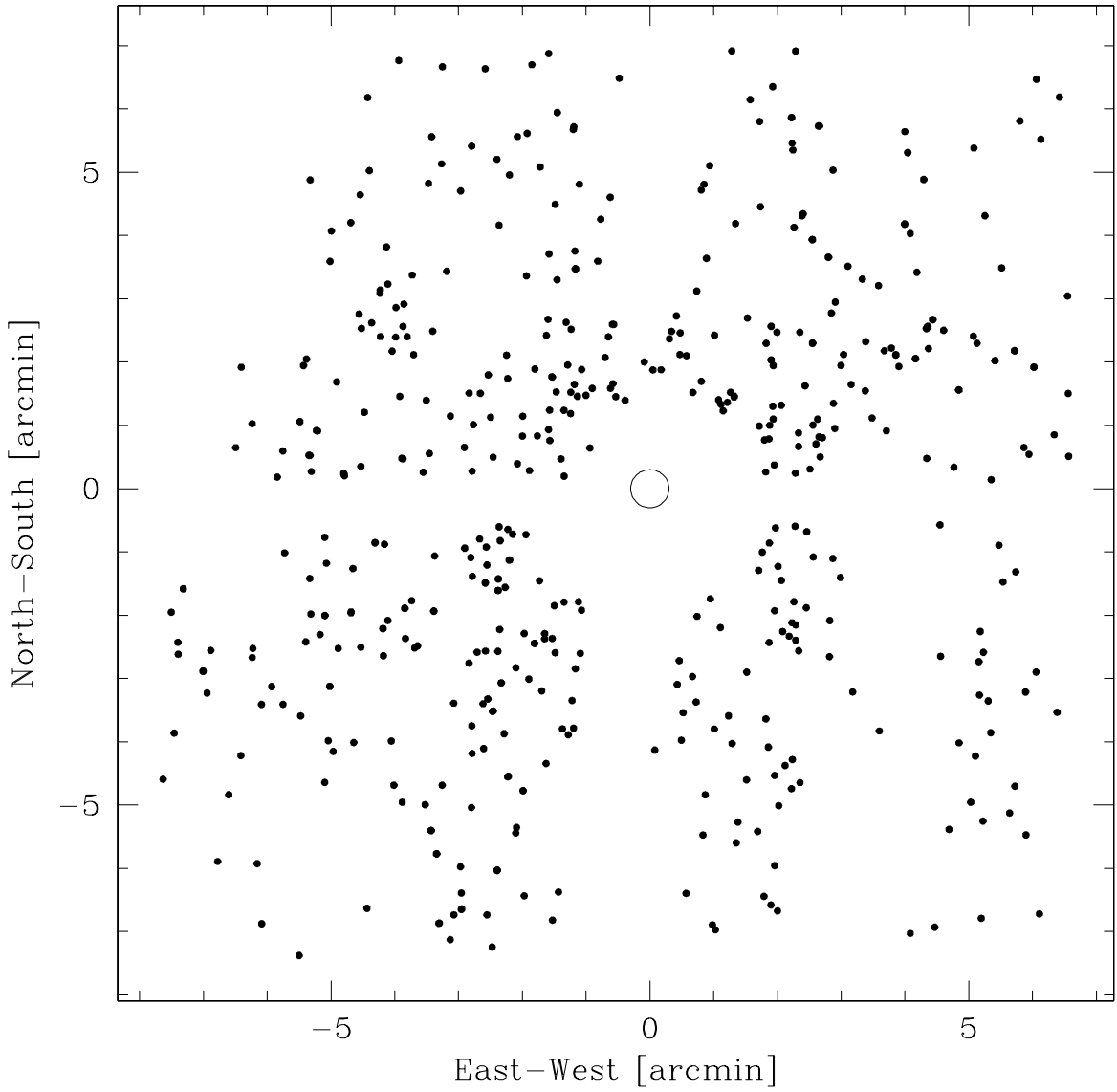}}}
\caption{Distribution of the final cluster sample. The open circle denotes
	the center of NGC\,1399.}
\label{fig:cand_distr}
\end{figure}

\begin{figure}[t]
\centerline{\resizebox{\hsize}{!}{\includegraphics{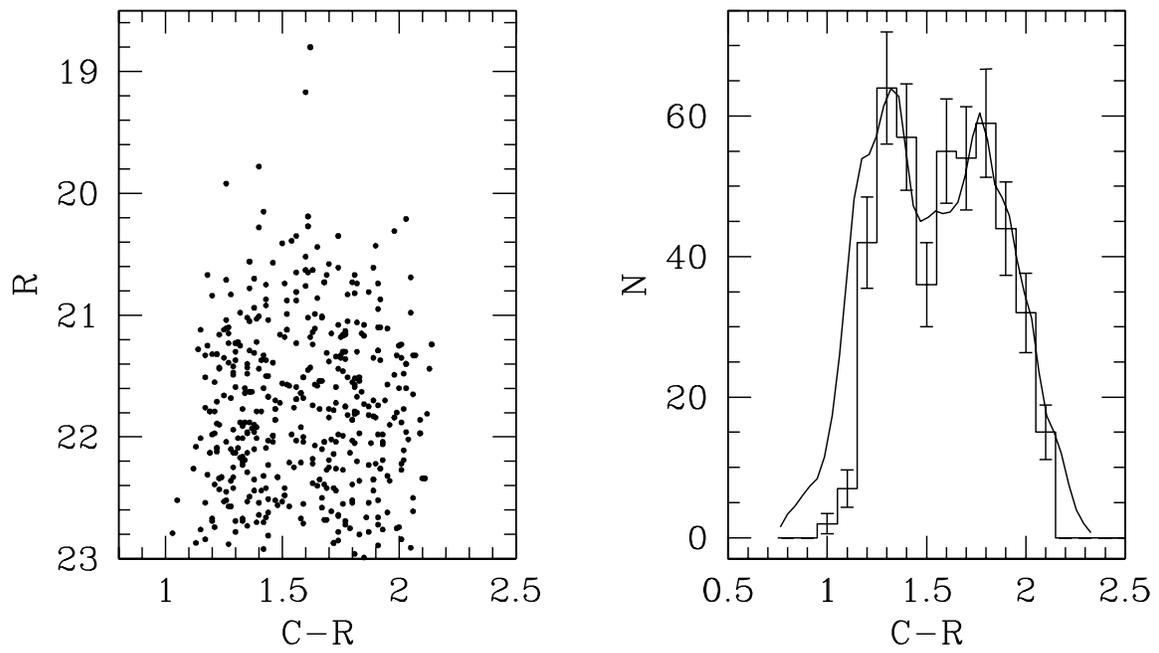}}}
\caption{{\bf Left panel:} color magnitude diagram of the final cluster sample.
	{\bf Right panel:} color distribution of the final sample (histogram) is
	compared to the arbitrarily scaled, adaptively smoothed
	(Epanachikov kernel) color distribution of the whole MOSAIC sample
	(Paper\,{\sc i}) within
	the common radial range. The agreement between the two samples is very good
	and indicates that the spectroscopic sample is not biased.}
\label{fig:spec_cmd}
\end{figure}

The whole dataset is tabulated in Table\,\ref{tab:finaltab}.  The first column
identifies the cluster: we used an identifier consisting of two 2-digit numbers, the
first indicating the mask and the second the aperture number on the mask which
was assigned during the extraction process. Column two and three are right
ascension and declination of the clusters (J2000).  The positions are based on
the USNO\,2.0 catalogue
\footnote{http://tdc-www.harvard.edu/software/catalogs/ua2.html} (Urban et al.
\cite{urban98}).  The color and magnitude information in columns 4 and 5 are
taken from Dirsch et al. (\cite{dirsch03}). The uncertainties do not include the
photometric calibration errors that are 0.03 in C-T1 and 0.02 in T1. The sixth
column gives the velocity determined with the correlation measurement and the
seventh column the velocity determined using direct line measurements. The
uncertainties given are those returned by the respective packages used. The last
column is reserved for comments. An identifier for a different cluster indicates
that this cluster has been observed independently on two masks and can be
identified via its number on the other mask. The spatial distribution of the
whole cluster sample tabulated in Tab.\ref{tab:finaltab} is shown in Fig.\,\ref{fig:cand_distr}.

Color and magnitude information are missing for some clusters. The reason is the
incomplete coverage of the field due to the undithered gaps in the MOSAIC image
or that they are located within saturated regions caused by bright nearby
objects. These candidates had been selected ''by eye'' after all object slits on
the mask had been set as described above. The velocity given is the mean
value of the two correlation measurements and the error a simple mean error.
We quote the mean velocity only for the first appearance of the cluster in the list.

\begin{figure}[t]
\centerline{\resizebox{\hsize}{!}{\includegraphics{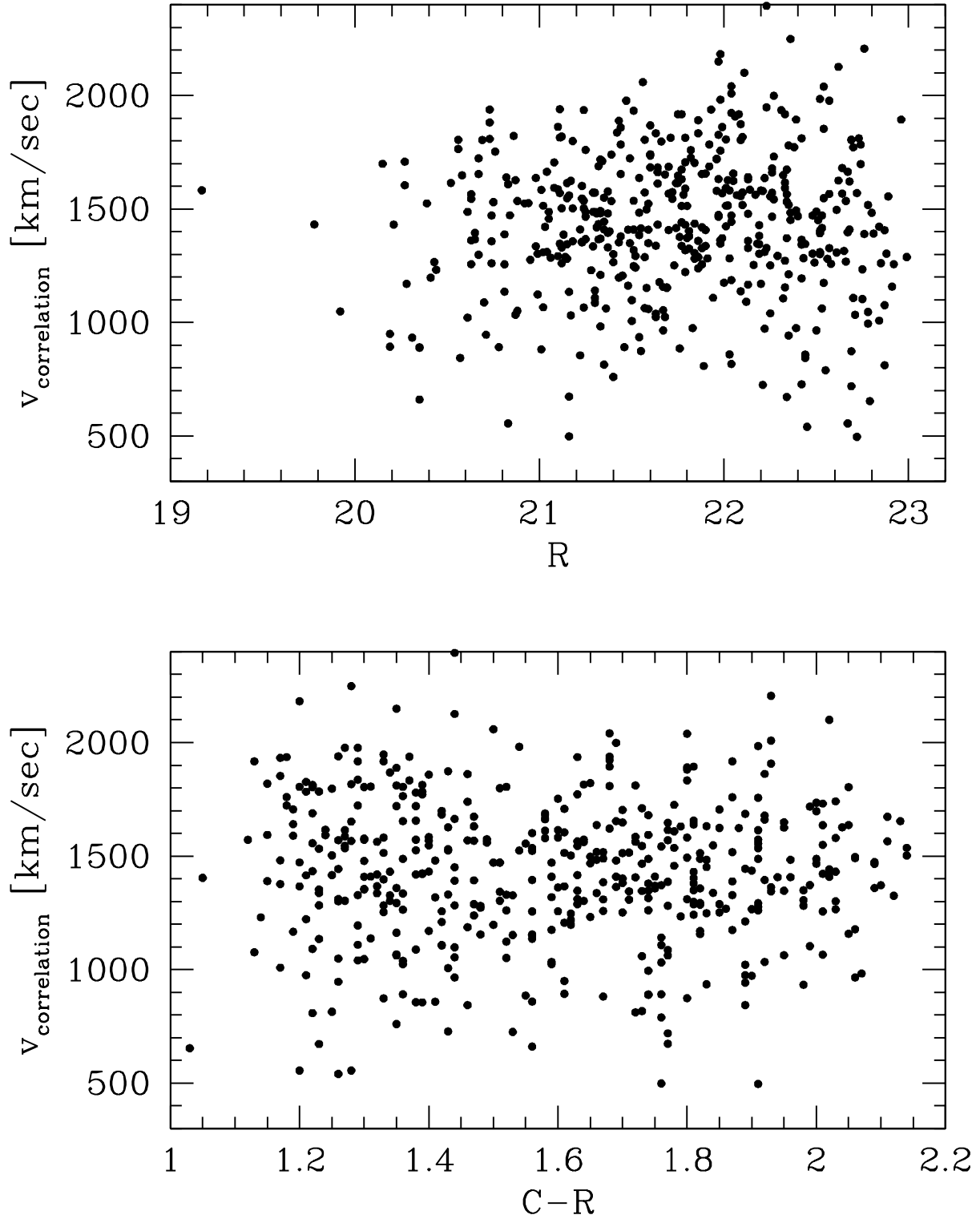}}}
\caption{Luminosity ({\bf upper panel}) and color ({\bf lower panel}) dependence
	of the correlation velocities. No correlation is seen with luminosity, however,
	the scatter of blue objects is larger than that of red objects, which is
	due to their different dynamics explained in Paper\,{\sc ii}.}
\label{fig:velo_rela}
\end{figure}

The color and magnitude dependence of the correlation velocities are shown in
Fig.\,\ref{fig:velo_rela}. The dynamical interpretation is given in
Paper\,{\sc ii}.

Stars in our sample for which velocities have been determined are tabulated in
Table\,4. In some cases no correlation velocities are given, only line velocities.
The reason is that our template is not particularly suited to be correlated with
late-type stars, which also explains the systematically larger uncertainties in the
correlation velocities.

\section{Comparison with the literature}

Globular clusters around NGC 1399 have been spectroscopically observed earlier.
The largest sample, containing 47 globular cluster velocities, has been obtained
by Grillmair et al. (\cite{grillmair94}) using the Anglo-Australian Telescope
with the Low-Dispersion Survey Spectrograph with $\approx13$\,\AA \ resolution,
which resulted in a velocity uncertainty of approximately 150\,km/sec.  Better
velocities with uncertainties of approximately 100\,km/sec have been measured for
18 clusters by Minniti et al. (\cite{minniti98}) with the ESO Multi-Mode
Instrument at the ESO New Technology Telescope with a resolution of 7.5\AA. The
observations done by Kissler-Patig et al. (\cite{kisslerpatig98}) with the Low
Resolution Imaging Spectrograph at Keck (resolution of 5.6\AA) yielded the best
velocities so far for, 18 globulars with uncertainties around 35\,km/sec. Some of
these globular clusters were observed in our run as well and in
Tab.\ref{tab:litcomp} our velocities are compared with the velocities determined
in the earlier studies. In addition we compiled stars that were observed in our
run and are also present in the Two Degree Field (2dF) Survey (Drinkwater et al.
\cite{drinkwater01}).

The mean difference and standard deviation between our velocities and those of
Kissler-Patig et al. (\cite{kisslerpatig98}) (we always subtract the reference
velocity from ours) is +101 km/sec and 58 km/sec, respectively. From the
published uncertainties we would expect a standard deviation of approximately 60
km/sec, which agrees well.  From the Grillmair et al. (\cite{grillmair94}) data
we find for the mean difference and standard deviation -293\,km/sec and
153\,km/sec, respectively. This large difference is driven mainly by a few
extremely deviating objects, probably caused by the low S/N of Grillmair et al.'s
spectra. However, also in this case the observed and expected standard deviations
are in good agreement. Regarding the stars in common with the 2dF survey, we find
a difference of -56\,km/sec with a standard deviation of 83\,km/sec, while we
would expect a standard deviation of 61\,km/sec. The reason for the slight
discrepancy probably is that we used line instead of correlation velocities for
the stars and we have shown earlier that the errors for the velocities obtained
via line measurements is underestimated. These measurements are hence in good
agreement with our velocities.

\section*{Acknowledgments}
BD, TR, DG, LI and DM gratefully acknowledge support from the Chilean Center for Astrophysics
FONDAP No. 15010003. BD gratefully acknowledges financial support of the
Alexander-von-Humboldt Foundation via a Feodor Lynen Stipendium. We thank our anonymous
referee for her/his helpful comments.

\clearpage


\begin{deluxetable}{cccc}
\tablecaption{Observed masks, their central positions and exposure times}
\tabletypesize{\scriptsize}
\tablewidth{0pt}
\tablehead{
	\colhead{field} &
	\colhead{mask \#} &
	\colhead{center position}&
	\colhead{$\mathrm{t}_\mathrm{exp}$}
}
\startdata
1&	75& 3:38:54 -35:23:30&90\,min\\
1&	76& 3:38:45 -35:23:30&45\,min 	\\
1&	77& 3:38:35 -35:23:27&45\,min\\
1&	78& 3:38:45 -35:25:08&90\,min\\
2&	80& 3:38:21 -35:23:30&90\,min\\
2&	81& 3:38:08 -35:23:31&45\,min\\
2&	82& 3:38:11 -35:25:25&45\,min\\
3&	84& 3:38:54 -35:31:00&45\,min\\
3&	86& 3:38:37 -35:30:59&90\,min\\
4&	89& 3:38:17 -35:31:01&45\,min\\
4&	90& 3:38:13 -35:31:01&90\,min\\
4&	91& 3:38:01 -35:31:01&90\,min\\
4&	92& 3:38:08 -35:29:16&45\,min\\
\enddata
\label{tab:observations}
\end{deluxetable}

\begin{deluxetable}{cccccc}
\tablecaption{Summary of number and type of observed objects on each mask.
Col.(2), (3), (4) give the number of objects for which we obtained a spectrum.
Col.(6) is the number of clusters for which a velocity has been determined. }
\tabletypesize{\scriptsize}
\tablewidth{0pt}
\tablehead{
        \colhead{mask \#} &
        \colhead{candidates} &
        \colhead{other point} &
        \colhead{galaxies} &
	\colhead{sky} &
	\colhead{obtained} \\
	\colhead{}&
	\colhead{} &
	\colhead{sources} &
	\colhead{} &
	\colhead{slits} &
	\colhead{velocities}
}
\startdata
75& 32 & 11 & 16 & 51 & 27  \\
76& 45 & 12 & 16 & 47 & 40  \\
77& 42 & 14 &  6 & 59 & 37  \\
78& 32 & 29 & 10 & 45 & 37  \\
80& 38 & 19 &  4 & 54 & 49  \\
81& 39 & 15 & 16 & 37 & 42  \\
82& 46 &  5 &  8 & 27 & 36  \\
84& 35 &  6 & 31 & 34 & 26  \\
86& 61 &  4 &  6 & 39 & 59  \\
89& 50 & 13 & 13 & 28 & 48  \\
90& 28 & 28 & 14 & 34 & 43  \\
91& 32 & 17 & 16 & 44 & 29  \\
92& 22 & 16 & 19 & 40 & 29  \\\hline
total& 502 & 189 & 175 & 539 & 502 \\
\enddata
\label{tab:efficiency}
\end{deluxetable}

\clearpage

\begin{deluxetable}{cccccccc}
\tabletypesize{\scriptsize}
\tablecaption{Globular clusters for which velocities have been determined with
	lines and/or cross correlation. The coordinates are tied to the USNO-A2.0
	catalogue. The appearance of a cluster identifier in the last column indicates
	that two velocities have been obtained independently on two masks. In this
	case the velocity, in the last column, is the mean value from the two masks. }
\tablewidth{0pt}
\tablehead{
	\colhead{Identifier}&
	\colhead{R.A. (2000)} &
	\colhead{Decl. (2000)} &
	\colhead{C-R}&
	\colhead{R}&
	\colhead{v$_c$ [km/sec]}&
	\colhead{v$_l$ [km/sec]}&
	\colhead{comments}
}
\startdata
\input{dirsch.tab1.dat}
\enddata
\label{tab:finaltab}
\end{deluxetable}

\begin{deluxetable}{cccccccc}
\tabletypesize{\scriptsize}
\tablecaption{Stars in our sample for which velocities have been determined with
	lines or cross correlation. The meaning of the columns is the same as in
	Tab.\,\ref{tab:finaltab}.}
\tablewidth{0pt}
\tablehead{
	\colhead{Identifier}&
	\colhead{R.A. (2000)} &
	\colhead{Decl. (2000)} &
	\colhead{C-R}&
	\colhead{R}&
	\colhead{v$_c$ [km/sec]}&
	\colhead{v$_l$ [km/sec]}&
	\colhead{comments}
}
\startdata
\input{dirsch.tab2.dat}
\enddata
\label{tab:startab}
\end{deluxetable}

\clearpage

\begin{deluxetable}{ccccc}
\tabletypesize{\scriptsize}
\tablecaption{Compilation of clusters and stars for which
velocities have
previously been published in the literature. The nomination
refers to: {\it keck} - Kissler-Patig et al. (\cite{kisslerpatig98}),
{\it ntt} -- Minnitti et al. (\cite{minniti98}),
{\it aat} -- Grillmair et al. (\cite{grillmair94}, {\it 2dF} --
Drinkwater et al. (\cite{drinkwater01a}).}
\tablewidth{0pt}
\tablehead{
	\colhead{object} &
	\colhead{our velocity [km/sec]}&
	\colhead{other velocity [km/sec]} &
	\colhead{other denomination}
}
\startdata
78:115&$ 1358 \pm 34 $& $ 1150 \pm  59$ & keck 19 \\
80:19 &$ 1266 \pm 30 $& $ 1260 \pm  66$ & keck 14 \\
80:28 &$ 1753 \pm 18 $& $ 1688 \pm  42$ & keck 18 \\
80:35 &$ 1536 \pm 21 $& $ 1523 \pm  30$ & keck 15 \\
80:39 &$  881 \pm 18 $& $  866 \pm  42$ & keck 17 \\
80:57 &$ 1309 \pm 27 $& $ 1374 \pm  126$ & keck 20 \\ ??
82:51 &$ 1933 \pm 75 $& $ 1732 \pm  42$ & keck 12 \\
90:74 &$  946 \pm 20 $& $  732 \pm  32$ & keck 1 \\
\enddata
\label{tab:litcomp}
\end{deluxetable}

\begin{figure}[t]
\centerline{\resizebox{\hsize}{!}{\includegraphics{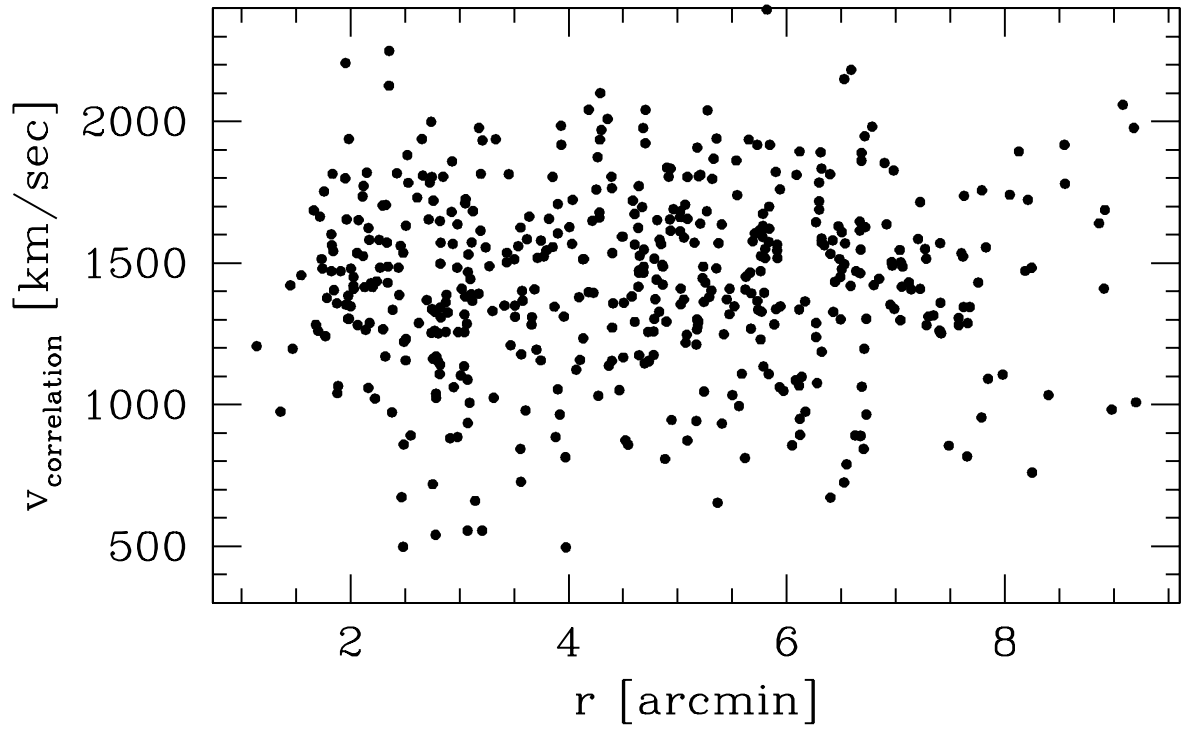}}}
\caption{The dependence of the correlation velocities on radial distance of NGC\,1399
	is shown. }
\label{fig:velo_radial}
\end{figure}

\end{document}

%% file: dirsch.tab1.dat
75:1  & 3:38:55.58 & -35:26:52.4 & $1.17 \pm 0.03$ & $21.33 \pm 0.01$ & $1481 \pm 52$& $1394 \pm 31$ &             \\ 
75:8  & 3:38:50.62 & -35:26:32.3 & $1.31 \pm 0.05$ & $22.09 \pm 0.03$ & $1137 \pm 65$& $1254 \pm 29$ &             \\
75:9  & 3:39:01.54 & -35:26:30.1 & $1.38 \pm 0.03$ & $21.31 \pm 0.02$ & $1419 \pm 47$& $1938 \pm 22$ &             \\
75:24 & 3:39:01.50 & -35:25:30.6 & $1.51 \pm 0.09$ & $22.48 \pm 0.04$ & $1303 \pm 59$& $1400 \pm 54$ &             \\
75:25 & 3:38:53.08 & -35:25:27.3 & $1.30 \pm 0.08$ & $22.69 \pm 0.04$ & $1804 \pm 57$& $1708 \pm 41$ &             \\
75:31 & 3:38:58.86 & -35:25:05.6 & $1.38 \pm 0.03$ & $22.10 \pm 0.02$ & $1572 \pm 51$& $1563 \pm 31$ &     78:11 :: $1579 \pm 44$ \\
75:33 & 3:38:55.86 & -35:24:59.6 & $1.24 \pm 0.06$ & $22.33 \pm 0.04$ & $1616 \pm 38$& $1528 \pm 32$ &     75:33 :: $1604 \pm 39$ \\
75:34 & 3:38:49.74 & -35:24:57.6 & $2.09 \pm 0.08$ & $21.97 \pm 0.04$ & $1473 \pm 42$& $1460 \pm 25$ &     78:43 :: $1469 \pm 42$ \\

%% file: dirsch.tab2.dat
75:15&  3:38:59.48& -35:26:05.9&                   &                   &$  120 \pm 140$ &               & 78:9\\ 
75:53&  3:38:55.31& -35:23:47.9&                   &                   &$  114 \pm  67$ &$  42 \pm 18 $ &\\ 
75:55&  3:38:58.17& -35:23:39.8&                   &                   &$   71 \pm  61$ &$  -4 \pm 23 $ &\\ 
75:57&  3:38:57.87& -35:23:32.5&                   &                   &$  -77 \pm  90$ &$ -51 \pm 31 $ &\\ 
75:83&  3:38:50.48& -35:22:13.4&$  1.78 \pm 0.03 $ &$ 20.83 \pm 0.02 $ &$  145 \pm  95$ &$            $ &\\ 
75:87&  3:38:55.78& -35:21:59.4&$  0.88 \pm 0.01 $ &$ 17.93 \pm 0.01 $ &$  168 \pm  26$ &$ -31 \pm 22 $ &\\
75:89&  3:38:51.42& -35:21:51.9&$  2.41 \pm 0.02 $ &$ 17.86 \pm 0.01 $ &$   32 \pm  39$ &$ 153 \pm 15 $ &\\
75:90&  3:38:57.66& -35:21:47.1&$  3.12 \pm 0.04 $ &$ 19.81 \pm 0.02 $ &$   39 \pm  79$ &$ -77 \pm 33 $ &\\